\documentclass[sigplan,screen]{cidr-2025}
\usepackage{enumitem}

\newcommand{\textred}[1]{\textcolor{red}{#1}}
\ifx\noeditingmarks\undefined
  \newcommand{\pgwrapper}[2]{\textred{#1: }\textit{#2}}
\else
  \newcommand{\pgwrapper}[2]{}
\fi

\AtBeginDocument{%
  }

\begin{document}

\title{Bullion: A Column Store for Machine Learning}



\author{Gang Liao}
\authornote{This work was inspired in Gang's last quarter at ByteDance.}
\email{gangliao@umd.edu}
\affiliation{%
  \institution{University of Maryland}
  \city{Collge Park}
    \state{MD}
  \country{USA}}

\author{Ye Liu}
\email{ye.liu@bytedance.com}
\affiliation{%
  \institution{ByteDance Infra System Lab}
  \city{San Jose}
    \state{CA}
  \country{USA}}

\author{Jianjun Chen}
\email{jianjun.chen@bytedance.com}
\affiliation{%
  \institution{ByteDance Infra System Lab}
  \city{San Jose}
    \state{CA}
  \country{USA}}

\author{Daniel J. Abadi}
\email{abadi@cs.umd.edu}
\affiliation{%
  \institution{University of Maryland}
  \city{Collge Park}
    \state{MD}
  \country{USA}}

\renewcommand{\shortauthors}{Gang Liao et al.}

\begin{abstract}


The past two decades have witnessed significant success in applying columnar storage to data warehousing and analytics. However, the rapid growth of machine learning poses new challenges. This paper presents Bullion, a columnar storage system tailored for machine learning workloads. Bullion addresses the complexities of data compliance, optimizes the encoding of long sequence sparse features, efficiently manages wide-table projections, introduces feature quantization in storage, enables quality-aware sequential reads for multimodal training data, and provides a comprehensive cascading encoding framework that unifies diverse encoding schemes through modular, composable interfaces. 
By aligning with the evolving requirements of ML applications, Bullion facilitates the application of columnar storage and processing to modern application scenarios such as those within advertising, recommendation systems, and Generative AI. 

Preliminary experimental results and theoretical analysis demonstrate Bullion's improved ability 
 to deliver strong performance in the face of the unique demands of machine learning workloads compared to existing columnar storage solutions. Bullion significantly reduces I/O costs for deletion compliance, achieves substantial storage savings with its optimized encoding scheme for sparse features, and improves metadata parsing speed for wide-table projections. These advancements enable Bullion to become an important component in the future of machine learning infrastructure, enabling organizations to efficiently manage and process the massive volumes of data required for training and inference in modern AI applications.
 
\end{abstract}





\maketitle

\section{Introduction}
\label{sec:intro}

Columnar storage, with its organization of data into individual columns, has reshaped data warehousing and big data analytics over the past two decades, offering many benefits, including optimized attribute skipping, effective data compression, and vectorized query processing~\cite{abadi2013design,abadi2006integrating,abadi2008column,abadi2009column,abadi2007column,ailamaki2001weaving}. 
In the early 2010s, 
open-source column-oriented file formats started to emerge, 
following the success of column-stores in academia and industry~\cite{stonebraker2018cstore,lamb2012vertica, nes2012monetdb,zukowski2012vectorwise}.
These open-source columnar formats~\cite{he2011rcfile, huai2014major, parquet, orc} have become the de facto standards for data warehouses, data lakes and lakehouses stored in the Hadoop and Spark ecosystems. 
Apache Parquet~\cite{parquet} and Apache ORC~\cite{orc} have become popular storage choices for data stored in Hadoop HDFS~\cite{hdfs,gang2023filescale} and Amazon S3~\cite{s3}. Their adoption extends to major data analytics frameworks, including Hive~\cite{hive, thusoo2009hive, thusoo2010hive}, Impala~\cite{bittorf2015impala}, Presto/Trino~\cite{sethi2019presto, prestodb, trino}, Dremio~\cite{dremio}, Flink~\cite{carbone2015apache}, Spark~\cite{spark} and Arrow DataFusion~\cite{datafusion,liao2024flock,gang_thesis}. Furthermore, even database products with proprietary storage formats, such as Redshift~\cite{redshift, gupta2015amazon, pandis2021evolution, armenatzoglou2022amazon}, Snowflake~\cite{dageville2016snowflake,snowflake}, ClickHouse~\cite{clickhouse}, and Google BigQuery~\cite{bigquery}), have extended their support to Parquet and ORC via external tables. The recent lakehouse~\cite{armbrust2021lakehouse} trend has further expanded the scope of these formats, emphasizing better metadata management (e.g., ACID transactions). Specifically, Delta Lake~\cite{armbrust2020delta}, Apache Iceberg~\cite{iceberg}, and Apache Hudi~\cite{hudi} are widely used  metadata management tools in this space that retain the columnar file structures of Parquet and ORC.


Machine learning workloads, often involving complex last-mile data transformations~\cite{audibert2023tf,murray2021tf} and feature engineering~\cite{anderson2016input,feature-engineering}, derive substantial benefits from the efficiency of column stores in data retrieval and query execution. This read efficiency is important for optimizing data pipelines, mitigating the limitations of training accelerators (e.g., GPUs and TPUs) operating under fixed power budgets in data centers, and thereby accelerating model training and experimentation~\cite{data-ingestion}. Furthermore, the capability of column stores to selectively process data significantly facilitates feature selection and extraction, empowering practitioners to efficiently distill relevant features from extensive training datasets, even those with over 20,000 columns~\cite{zhao2022understanding}.  Additionally, the improved support for schema evolution~\cite{rae2013online, bhattacherjee2021bullfrog} in column stores is important for machine learning workflows, accommodating the dynamic nature of data sources and facilitating the management of multiple versions of training datasets.  However, despite the advantages of columnar storage in applied machine learning, these formats, developed over a decade ago and rooted in early 2000s DBMS methodologies~\cite{abadi2013design}, have limitations that are becoming increasingly evident~\cite{zeng2023vldb}.  


%


{\bf Deletion Compliance}: In major internet companies, personalization and recommendation systems rely heavily on gathering and processing vast amounts of user data to create detailed user profiles. This data is transformed into training datasets for ML models, 
which are currently deployed for a variety of tasks including ad click-through rate (CTR) prediction and rankings. However, the tech industry faces the challenge of adhering to stringent AI privacy and diverse data compliance regulations worldwide. Laws such as the EU's GDPR~\cite{gdpr}, California's CCPA~\cite{ccpa}, and CPRA~\cite{cpra}, and Virginia's VCDPA~\cite{vcdpa}, mandate the physical deletion of user data within specified timeframes. This includes information from users who have canceled their accounts, opted out, or are underage. These deletions can be slow in columnar storage for two reasons. First, since each column in the deleted row is stored separately, a single request to delete a row results in many separate modifications: one for each column in that row. Furthermore, block-based compression is typically used, complicating direct modifications of individual rows.

{\bf Sparse Features and Wide Tables}: The diversity of features stored in columnar files and utilized by training jobs is substantial, as evidenced by an example Parquet file from ByteDance's ads table (see Table~\ref{tab:data_types}). 
This table contains over 16,256 columns of the 
\texttt{list<int64>} type alone, in addition to many other columns of other types. 
These sparse features exhibit unique sliding window patterns, presenting opportunities for optimized encoding schemes beyond those available today in open-source column-store file formats.

This table is a good example of production datasets at ByteDance, which are characterized by frequent introduction and deprecation of features, with several hundred modifications occurring monthly. 
They contain a broad spectrum of features, including those in beta, experimental, active, and deprecated stages, leading to a typical feature count that approaches approximately 20,000 columns. Machine learning training often requires reading only a small portion (about 10\%) of these columns~\cite{zhao2022understanding}, motivating the use of columnar storage.
However, this scenario introduces significant challenges, particularly in terms of metadata overhead and read efficiency, potentially causing performance bottlenecks in data retrieval and processing for machine learning workloads. 


\begin{table}[h!]
\centering
\begin{tabular}{|l|r|}
\hline
\textbf{Column Type} & \textbf{\# Columns} \\
\hline
\texttt{list<int64>} & {\bf 16,256} \\
\hline
\texttt{list<float>} & 812 \\
\hline
\texttt{list<list<int64>{>}} & 277 \\
\hline
\texttt{struct<list<int64>, list<float>{>}} & 143 \\
\hline
\texttt{struct<list<int64>{>}} & 120 \\
\hline
\texttt{struct<list<binary>{>}} & 46 \\ \hline
\texttt{struct<list<float>{>}} & 29 \\ \hline
\texttt{struct<list<binary>, list<binary>{>}} & 18 \\ \hline
\texttt{struct<list<double>{>}} & 10 \\ \hline
\texttt{list<binary>} & 8 \\  \hline
\texttt{struct<list<list<int64>{>}{>}} & 5 \\ \hline
\texttt{struct<list<binary>, list<float>{>}} & 5 \\ \hline
\texttt{string} & 3 \\ \hline
\texttt{int64} & 1 \\
\hline
\end{tabular}
\caption{Statistical breakdown of column types in an Ad Parquet file.}

\label{tab:data_types}
\end{table}

{\bf LLM Storage}: The proliferation of floating-point and embedding features in machine learning applications, particularly in large language models (LLMs), introduces significant challenges for data management systems. We now describe two important use cases that substantially impact columnar storage and batch processing requirements. 

{\bf 1) LLM training}: Pre-training involves the ingestion and curation of massive datasets from web crawls (e.g., Common Crawl~\cite{commoncrawl}), encompassing diverse data types including text, image, audio, and video.  Beyond traditional SQL operations (e.g., filters and joins), modern data pipelines require extensive preprocessing workflows~\cite{llama} incorporating offline batch inference across multiple ML models. This preprocessing encompasses data normalization, resizing, integrity and privacy filtering, quality scoring, content extraction (e.g., OCR for video frames), and caption auto-generation with subsequent scoring. The resultant feature backfilling process generates numerous derived columns-- including embeddings--which significantly amplifies both the cardinality and dimensionality of the underlying data tables.

Multimodal training data poses unique storage challenges due to its unstructured nature and heterogeneous sources. To efficiently manage such diverse data types, we adopt a hybrid storage architecture: leveraging columnar storage for structured metadata and embeddings, while utilizing Avro~\cite{avro}--a row-oriented binary format with schema support--for chunked storage of large media objects (e.g., video and audio content). However, this hybrid architecture introduces non-trivial I/O challenges during training. While metadata and quality scores are maintained in columnar storage (meta tables) and the corresponding media content in binary format (media tables), the access of high-quality samples via filtering criteria (e.g., quality scores) selected for training require bouncing back and forth across 
both meta and media tables. This scattered data layout leads to random I/O patterns, potentially decreasing training throughput.
Furthermore, the contextual relationships between text, audio, image, and video data, combined with privacy encryption requirements, present additional technical challenges in effective columnar storage design.

{\bf 2) LLM serving}: with the emergence of LLM-powered AI search engines~\cite{perplexity, searchgpt}, retrieval-augmented generation (RAG) has become a critical paradigm for managing diverse unstructured data--ranging from images and audio to text--through vector embeddings. These embeddings, retrieved via vector similarity search during inference, enrich prompts by incorporating contextual information, historical data, and relevant knowledge. 
These systems must crawl and analyze over 10 billion web pages daily, performing large-scale offline batch inference to generate trillions of embeddings over time. While these embeddings enable efficient retrieval and LLM-based summarization of relevant content for user queries, they also demand near-real-time data ingestion and processing capabilities, particularly for time-sensitive domain-specific sources (e.g., sports, news, weather) at hourly or minute-level granularity. The 
challenges manifest in two dimensions: managing 
columnar storage at this scale,
and maintaining low-latency embedding generation and retrieval to support real-time inference.

\begin{figure}
\centering
\includegraphics[width=0.5\textwidth]{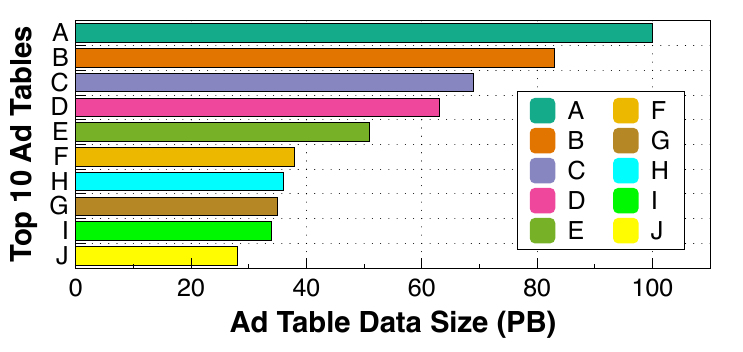}
\caption{Top 10 Ad tables in CN region.}
\label{top10}
\end{figure}

The exponential growth in industrial AI systems, particularly recommendation models and LLMs, is fundamentally constrained by computational and storage resource limitations that affect production deployment costs and scalability. The storage constraints are especially pronounced in large-scale advertising infrastructures - individual tables in ByteDance's CN region can approach 100PB in size, as shown in Figure~\ref{top10}. 
Other major technology companies including Meta Platforms~\cite{data-ingestion}\cite{zhao2022understanding}\cite{Chattopadhyay2023SharedFM} have reported similar scalability issues in their production environments. As model architectures continue to expand in both parameter count and feature dimensionality, their resource requirements grow commensurately. These 
observations demonstrate the need for innovative approaches to data management systems that can efficiently handle the increasing scale and complexity of modern ML workloads.

In this paper, we present Bullion, a next generation columnar storage system that addresses the limitations of legacy formats and 
accelerates modern advanced analytics and machine learning workloads. Its key contributions include: 1) a hybrid approach for deletion-compliant data removal, 2) a tailored delta encoding scheme for long sequence sparse features, 3) a compact, binary metadata layout for efficient wide-table projection and 4) the adaptation of model quantization techniques to feature quantization in storage, 5) a quality-aware data organization strategy for efficient multimodal training data access, and 6) a comprehensive cascading encoding framework with modular, composable interfaces for optimal compression. These advancements enable Bullion to meet the complex demands of modern data compliance, long sequence feature encoding, storage quantization, and wide table management, multimodal data handling, and adaptive compression, ultimately enhancing the efficiency and performance of large-scale training and inference processes in AI and machine learning contexts.

\section{New Challenges and Opportunities}

This section discusses the emerging challenges and potential opportunities facing columnar storage in AI contexts, and describes how Bullion is designed to address these challenges and take advantage of these opportunities. 
We explore six key areas: deletion-compliant data management in response to evolving privacy regulations, efficient encoding of sparse features with long sequences, optimized wide table management for feature-rich datasets, storage quantization for embedding data, quality-aware multimodal data organization for efficient LLM training, and a unified cascading encoding framework for optimal compression. These advancements aim to meet regulatory demands, reduce storage costs, and significantly enhance the efficiency of large-scale training and inference processes in machine learning workloads. 

\subsection{Deletion-Compliance}

Without optimizations, deleting a single row in columnar storage may require rewriting the entire file, leading to significant I/O and resource consumption.  Therefore, out-of-place paradigms are often used.
For example, Databricks, via their Delta Lake solution, utilizes deletion vectors~\cite{deletionvec} that employs bitmaps to mark rows for deletion. 
ByteHTAP~\cite{chen2022bytehtap} adopts a similar strategy to enhance garbage collection and merge-on-read processes. These methods enable marking changes without rewriting entire files, with subsequent reads merging these markers into the data being read.
However, this approach, which marks data as invalid, often does not align with data compliance regulations requiring timely deletion~\cite{mano2022deletion}. The delay in permanent data removal, where data remains in existence in storage despite being invisible via 
user requests, conflicts with the stringent timelines mandated by privacy laws. Therefore, in practice, many organizations avoid using these methods, and instead immediately delete data despite the cost. This has led to delete requests causing rewriting of hundreds of petabytes per month at TikTok, despite the fact that only 5\% of each file contains non-compliant data.

{\bf \noindent Solution.} Bullion introduces a hybrid approach to tackle the challenge of compliant and timely data erasure. It performs in-place updates to physically remove data, yet also uses deletion vectors to efficiently indicate which rows have had this update performed to them, so that they can be skipped during query execution. This approach ensures regulatory-compliant data removal without requiring full file reads /rewrites. Bullion uses metadata in the file footer to indicate which rows are marked for deletion, along with their row group and page offsets. This allows for direct, in-place updating of the associated pages in storage. This process must adhere to {\bf a key criterion}: the post-update page dimensions do not exceed their initial size, which is vital to upholding data integrity.

To achieve this, Bullion selectively deletes or masks data at its source, 
without interfering with how it is encoded.
Bullion supports a range of encoding methods, including: fixed-length bit-packing, dictionary encoding (Dict), variable-length integer (Varint), run-length encoding (RLE), and frame-of-reference (FOR-delta). Updates are made to encoded data while guaranteeing the size of the updated pages remain within their original limits, adhering to the  requirement of size consistency.

\begin{figure}
\centering
\includegraphics[width=0.5\textwidth]{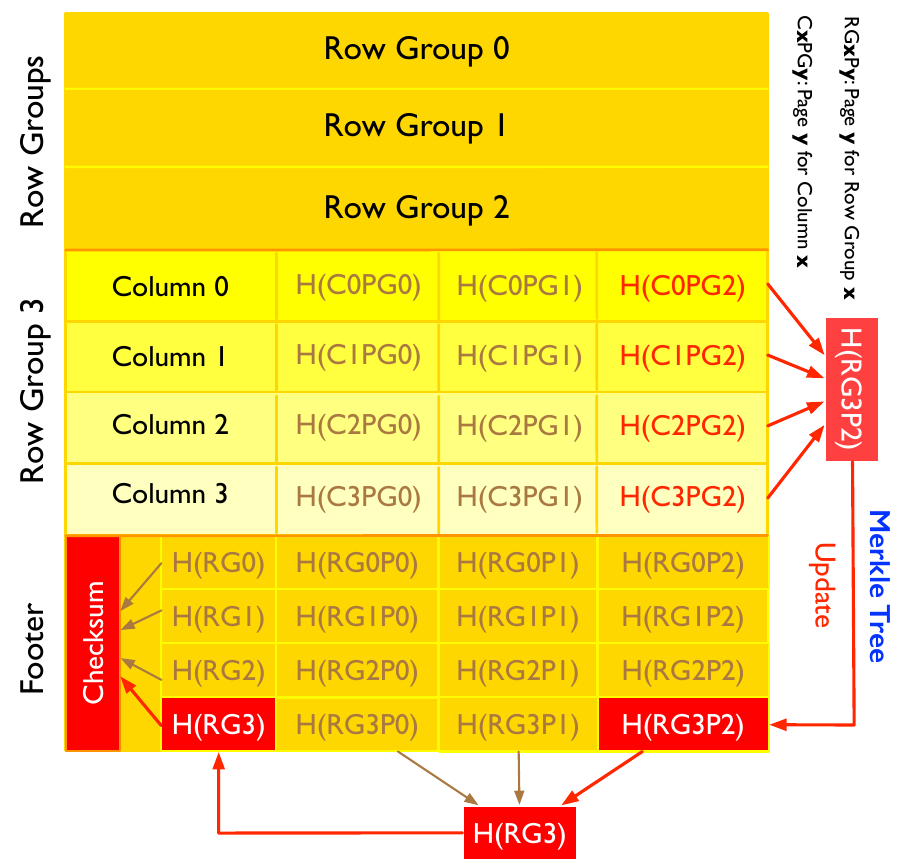}
\caption{Merkle tree update for checksum maintenance.}
\label{checksum}
\end{figure}

\noindent {\bf Bit-Packed Encoding} stores data elements using a minimal, fixed number of bits. 
Since the encoded values have a fixed size, it is straightforward to map bits in a bitmap to the encoded data elements, in order to mask deleted data. 

\noindent {\bf Varint Encoding} represents unsigned integers using the widely adopted LEB128 algorithm~\cite{leb128, liao2024sfvint}, encoding integers into byte sequences using less space for smaller values and a scaling mechanism for larger ones, where each byte holds 7 bits of the integer plus a continuation bit. 
When locating the encoded integer to be deleted, it suffices to retain the MSB (continuation bit) of each byte unchanged, while masking out the remaining 7 bits.

\noindent {\bf RLE Encoding} condenses sequences by representing consecutive identical elements with a single value and its occurrence count. 
In RLE, directly masking deleted elements is insufficient as it may lead to enlarged data post-re-encoding. Consider the sequence \texttt{222666663} with '3' occurrences of '2', '5' of '6', and '1' of '3', initially encoded as \texttt{236531}. Deleting the third '6' and masking it as '0' results in \texttt{222660663}, re-encoded as \texttt{2362016231}\footnote{For simplicity, the same method is used for both the re-encoding and initial encoding processes.} which is larger than the original encoding
. Instead, a deletion vector can be used, which details the valid values and their offsets in a page.
This keeps the data compact and aligned, as misaligned values are restored using the deletion vector.  For example, the same original data (\texttt{222666663}) becomes \texttt{22266663} after deletion, which is encoded to \texttt{236431} and a deletion vector \texttt{000001000}. This vector in the file footer allows \texttt{236431} to be decoded into \texttt{22266\textbf{X}663}.

\noindent {\bf Dictionary Encoding} constructs a dictionary to replace long, variable-length values of a specific domain to shorter (fixed-length) integer codes. 
This dictionary is stored in the file footer, which remains unmodified during the deletion process. Bullion introduces a default mask value entry within the dictionary, enabling efficient deletion by simply updating the integer code in the data pages to reference this mask entry. This approach effectively masks out the deleted values while maintaining the integrity of the residual data. 
It also allows
the integer codes in the data pages to
be further compressed using encoding techniques such as RLE, using the approach described above.

\noindent {\bf FOR-delta Encoding} declares a base value for each block or references the preceding value in a sequence, encoding data as deltas relative to these values. It supports random access to any element, 
and is often coupled with bit-packing. Bullion's deletion technique remains effective even with nested encoding schemes such as BtrBlocks~\cite{kuschewski2023btrblocks}.

For all of these encoding schemes, deletes  occur by only rewriting the page in which the deleted row resides instead of the entire file. 
When deleting 2\% of rows within a file,
data rewrite I/O costs can decrease by up to a factor of 50.
Furthermore, storage costs are nearly halved when full file rewrites are eliminated.

In addition to updating the encoded data, the checksums also must be updated in order to maintain data integrity and accuracy. 
Bullion assigns distinctive hash values to each page within the columnar file, as shown in Figure~\ref{checksum}. These granular hash values form the foundation for the computation of higher-level checksums at the row group tier. Subsequently, these checksums coalesce to formulate the overall file checksum, akin to a Merkle tree.  This Merkle tree structure eschews the traditional, monolithic approach (typically used by the open columnar formats used today) of recalculating checksums for the entire file, favoring instead page-level updates. Updates necessitate the propagation of the modified node's new hash value through the tree's hierarchy, culminating at the root, as depicted by the red arrows in the figure.  This incremental update mechanism ensures that only file segments affected by the change are read, boosting the efficiency of the checksum maintenance process.  

Bullion's deletion-compliance mechanism is backwards compatible with existing columnar formats such as Parquet and ORC, offering a flexible, privacy-aware approach through configurable compliance levels. At Level 0, the system operates as standard Parquet or ORC, maintaining complete backward compatibility but without upgraded deletion support. Level 1 implements deletion vectors, enabling query-time filtering of deleted rows while preserving the original data structure, suitable for scenarios with moderate privacy requirements. Level 2 combines deletion vectors with in-place updates, providing the highest degree of privacy compliance by physically removing sensitive data while maintaining optimal query performance. This tiered approach allows organizations to balance privacy requirements, performance considerations, and regulatory compliance based on data sensitivity and operational needs. Systems can dynamically adjust these levels on a per-table or per-column basis, facilitating fine-grained privacy control while ensuring seamless integration with existing data processing pipelines.

\subsection{Sparse Features Encoding}

\begin{figure*}[!h]
\centering
\includegraphics[width=0.95\textwidth]{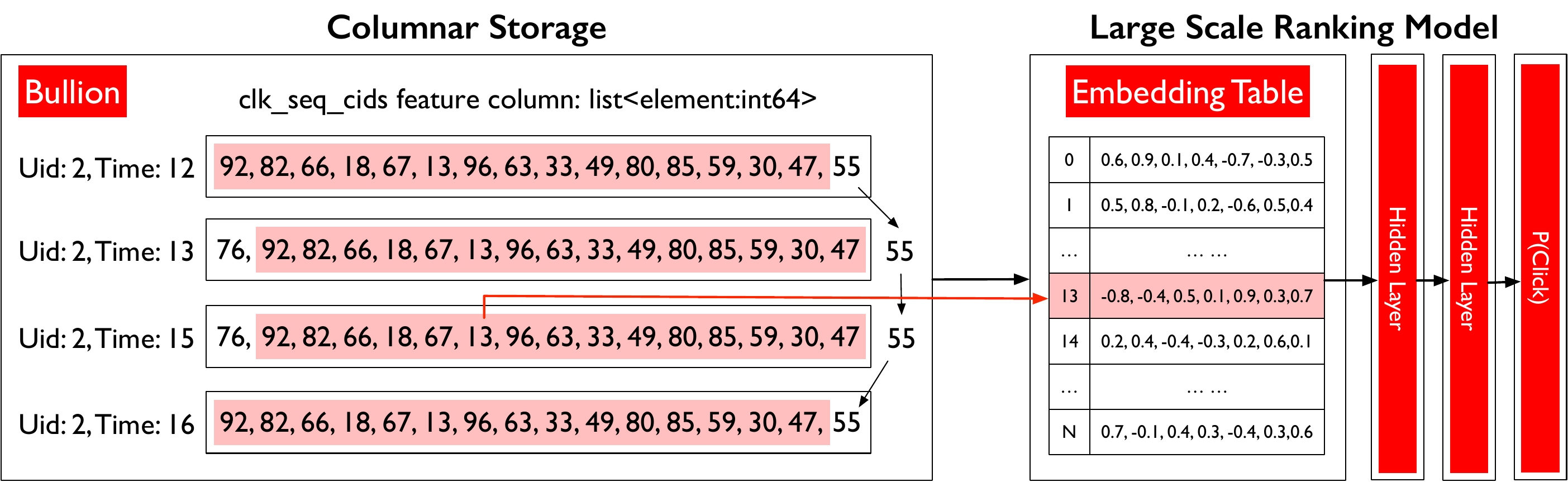}
\caption{Illustration of sliding window patterns in the simplified \texttt{clk\_seq\_cids} sparse feature column, represented as \texttt{list<element:int64>} data type, demonstrating temporal user engagement trends.}
\label{sliding_window}
\end{figure*}

Within recommendation systems, ranking is a key function, often powered by deep learning recommendation models. These models take a variety of inputs, including {\bf item features} such as keywords, image embeddings, related entity IDs, and historical click counts; {\bf user profiles} including age, gender, interests, and recent interactions; as well as {\bf contextual features} like city and phone type to yield a score in the output that ultimately decides the placement of an item, thus enabling efficient sorting and prioritization of content for display on websites and mobile applications. These ranking models primarily rely on sparse features, reflecting the reality that most users have not interacted with or rated the majority of items within the system. 
These sparse features (e.g., sets of IDs corresponding to information such as user interests) are transformed into float vectors suitable for neural network models. 


In columnar storage, it's common to store tens of thousands of sparse features (columns) for dynamic A/B testing in machine learning models. For example, \texttt{clk\_seq\_cids} is a feature represented as a vector of 256 int64 elements (stored as \texttt{list<int64>} in Parquet) where each element signifies an ad ID. This feature is essential for tracking user interactions with advertising campaigns over time, revealing patterns and trends in user engagement.  Typically, such data is categorized and sorted by user ID and timestamp before being written into columnar storage. As shown in Figure~\ref{sliding_window}, given the evolving nature of user interests over time, this sorting leads to the emergence of {\bf sliding window patterns} between vectors within the same feature column for individual users. These repetitive patterns present encoding optimization opportunities within the context of machine learning scenarios.

\begin{figure}
\centering
\includegraphics[width=0.5\textwidth]{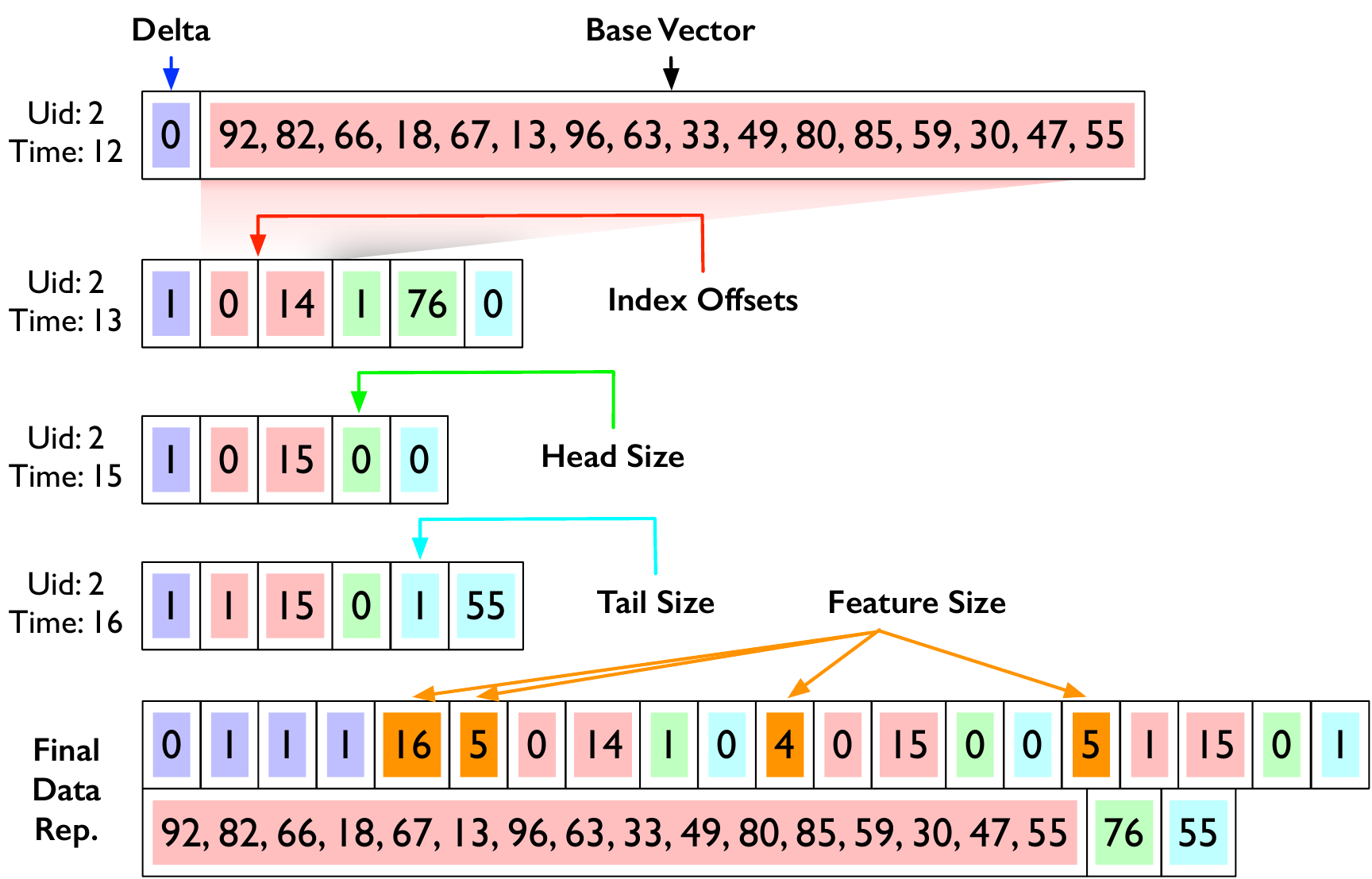}
\caption{Delta encoding for long sequence sparse features.}
\label{delta}
\end{figure}

{\bf Optimization.} Delta encoding is a good choice to encode datetime columns where data exhibits repetitive patterns in adjacent positions over time
since it needs to store only
the differences between successive values within the column.
However, current implementations of delta encoding in most columnar storage formats—whether open-source, such as Parquet and ORC, or proprietary formats used in cloud data warehouses—support only standard primitive types: \texttt{INT}, \texttt{BIGINT}, \texttt{DATE}, \texttt{TIMESTAMP}, and \texttt{DECIMAL}. 
Bullion extends delta encoding to long sequence vectors characteristic of sparse features in machine learning.
As shown in Figure~\ref{delta}, which illustrates the encoding process for the example provided in Figure~\ref{sliding_window}, the first vector of the column, \texttt{[92,82,66,18,67,...,85,59,30,47,55]}, serves as the base vector, using a delta flag set to 0 to denote the start of delta encoding. Subsequent feature encodings adopt the format: \texttt{<delta bit>} \texttt{<delta range>} \texttt{<len(head),data>} \texttt{<len(tail),data>}. For instance, the second feature differs from the base vector in the [0-14] range, with a new 76 at the head, and no new data at the tail. Conversely, the third feature is identical to the second, which is notated via the overlapping interval [0-15], with both head and tail sizes set to 0, indicating no new data and an unchanged sliding window. 
Figure~\ref{delta} shows the final data representation on disk: feature metadata and indexes are placed at the beginning, encoded via bitpacking or varint due to their smaller value. The bulk data follows, which can be compressed via zstd compression as machine learning training predominantly involves mini-batch reads with infrequent filtering. 

{\bf Challenge.} Traditional recommendation systems utilize labeled impressions as training data, where each record represents an atomic user-ad interaction. These training examples, derived from request logs, contain binary conversion labels indicating successful user actions. The volume of training data exhibits linear scaling with user engagement - a user with n ad impressions generates n distinct training records. However, recent advances in Generative Recommendation~\cite{chen2024hllm, zhai2024actions} mandate a paradigm shift from impression-centric to user-centric data modeling. This 
transition replaces discrete binary labels with temporal event sequences, where each user record encapsulates a comprehensive interaction history spanning both organic activities and advertising events (requests, impressions, and conversions). The evolution from simple binary supervision to complex temporal sequences poses significant challenges to existing infrastructure.

The current feature storage, training pipelines, and serving architectures, optimized for impression-based models, exhibit substantial limitations when handling these event sequences 
during both inference and training. 
This shift to user-centric data modeling necessitates a fundamental redesign of the storage stack. Users of existing columnar storage systems typically rely on suboptimal user-based bucketing and sorting mechanisms for data access, resulting in significant performance overhead when retrieving user event sequences. The storage system must evolve to support efficient representation and retrieval of long user event histories. Such evolution demands either substantial schema-level modifications or the development of novel storage formats that encapsulate rich temporal sequences of organic user events and advertising engagement events as a single training example per user. 

\subsection{Wide Table Projection}

As described in Section \ref{sec:intro}, the dynamic nature of production datasets is characterized by the frequent introduction and deprecation of features, with several hundred modifications occurring monthly~\cite{zhao2022understanding}. Columnar files may encompass a broad spectrum of features, including those in beta, experimental, active, and deprecated stages, leading to a feature count in the scale of tens of thousands of columns. Despite such extensive feature sets, individual training jobs necessitate the retrieval of only a specific subset of features, defined through a feature projection. This projection delineates the precise list of desired features for reading.
In practice, each training job may require access to less than 10\% of the stored features~\cite{zhao2022understanding}. 

For ML training datasets with over 10,000 columns, the overhead of reading metadata is nearly equivalent to the time needed to read 10\% of all columns. 
Thus, the time to read the metadata may double the read costs
(see for example
Figure 11 of Zeng et al.'s study on wide-table projection efficiency~\cite{zeng2023vldb}).

{\bf \noindent Solution.} Bullion adopts a compact metadata layout 
that enables direct metadata access from the footer, allowing for immediate buffer value reads {\bf without deserialization}. 
This binary format is reminiscent of Cap'n Proto~\cite{capnproto} and FlatBuffers~\cite{flatbuffer}. To access columns in Bullion files, the process begins with a \texttt{pread()} of the footer, followed by a binary map scan to find column indices. Byte ranges for each column are identified via an offsets array, followed by a targeted \texttt{pread()} for data retrieval.

{\footnotesize
\begin{verbatim}
table BullionFooter {
    num_rows: uint64;   page_compression_types: [uint8];
    rows_per_page:  [uint32];   page_offsets:   [uint64];
    pages_per_group:[uint32];   group_offsets:  [uint64];
    column_sizes:   [uint32];   column_offsets: [uint32];
    deletion_vec:   [uint64];   checksums:      [uint64];
    schema:         [Column];
}
\end{verbatim}
}

\begin{figure}
\centering
\includegraphics[width=0.5\textwidth]{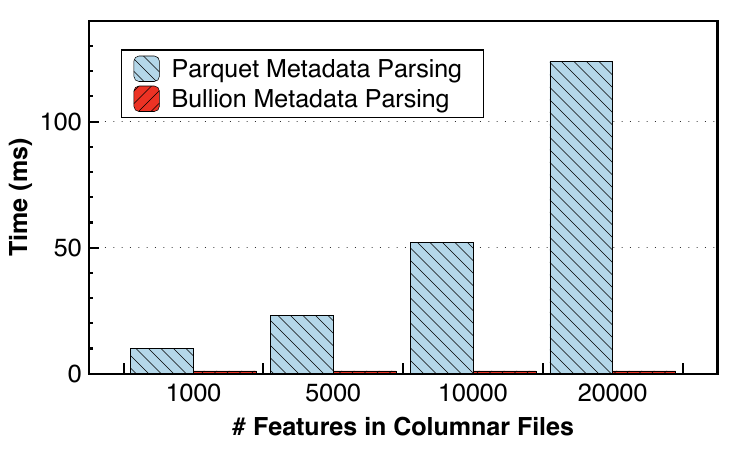}
\caption{Metadata parsing overhead in feature projection.}
\label{metadata_parsing}
\end{figure}

Figure~\ref{metadata_parsing} shows the time it takes Parquet and Bullion to extract a single column from a dataset consisting of a variable number of features. The figure shows that while Parquet's performance is significantly dependent on the number of feature columns, with retrieval time increasing linearly with the number of features, Bullion's performance is not (staying flat at less than 2 ms). When there are 10,000 feature columns, Parquet required 52 ms for metadata parsing, whereas Bullion required 1.2 ms.

This efficiency comes at the cost of flexibility, since the encoding format is fixed and lacks 
some customization options. 
This lack of flexibility is less problematic for machine learning tasks. since they often exhibit a simple data access pattern, reading all training data within a specific time period in a batch-oriented manner, 
without requiring complex indexing or filtering. 
Thus, the benefits of improved efficiency and faster data access outweigh the limitations in customization and extensibility. 


\subsection{Storage Quantization}

Feature and data representation 
are important determinants of model performance in both recommender systems and LLMs.  In recommender systems, the dimensionality of the feature space directly correlates with the size of embedding tables, which in turn determines the model's total memory footprint. However, production deployment scenarios frequently impose strict storage constraints that inhibit the incorporation of additional features and expanded embedding capabilities. As a consequence, numerous promising architectural improvements remain unrealized in production environments. 
This is particularly problematic for advertising platforms and AI search engines, where rich feature representations and embedding directly influence business outcomes.

\begin{figure}
\centering
\includegraphics[width=0.45\textwidth]{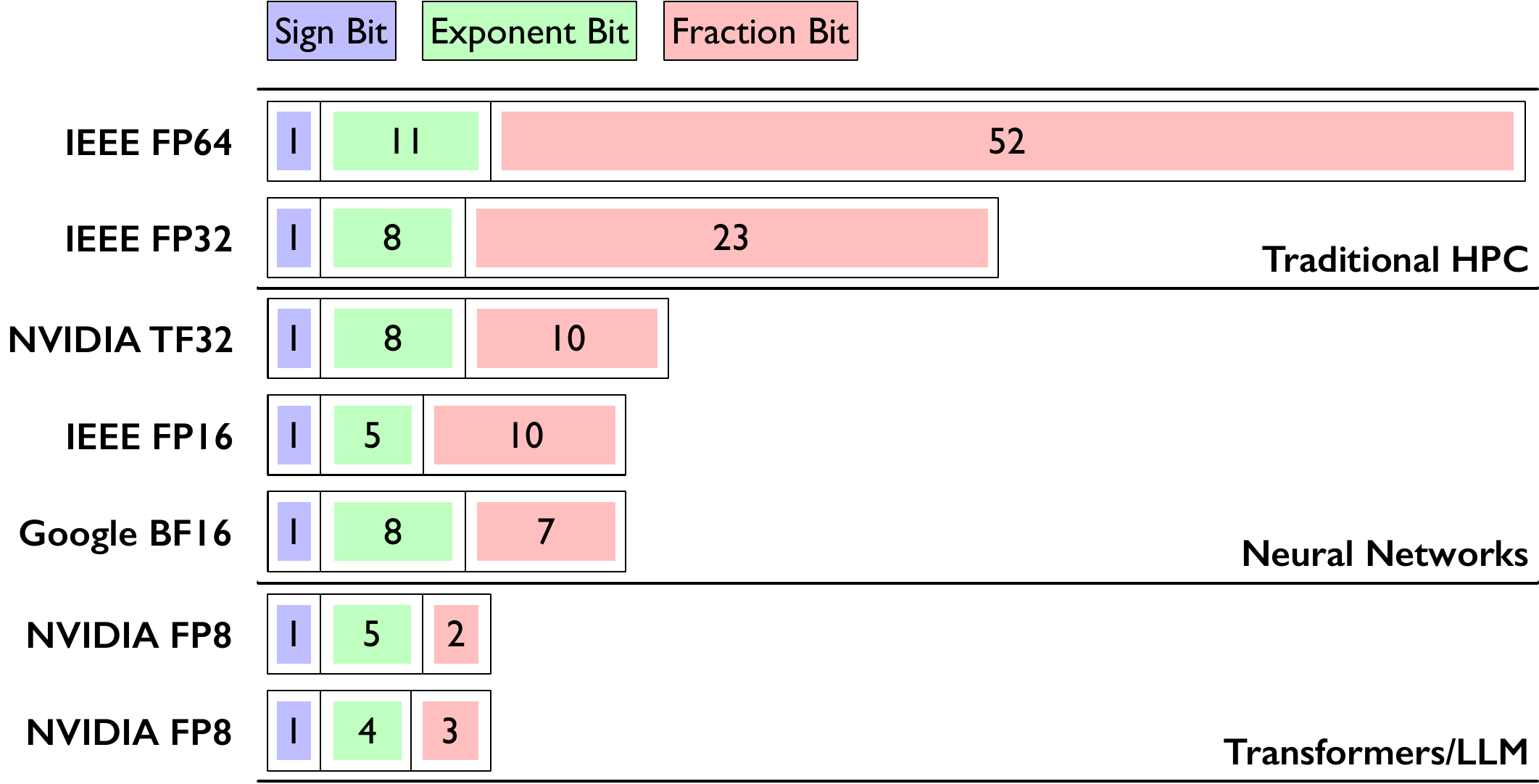}
\caption{Basics of floating–point quantization.}
\label{quant}
\end{figure}

Model quantization~\cite{han2015deep,jacob2018quantization,deng2020model}
aims to address model size, inference speed, and memory efficiency challenges. It typically reduces model weights from high-precision 32-bit floating-point (FP32) representations to more compact numerical formats, including lower-precision floating-point (FP16, FP8) or integer representations (see Figure~\ref{quant}). The proven effectiveness of model quantization in training and serving stages motivates its application for features and embeddings in storage, {\bf adapting model quantization to storage quantization}. 
Different features and embeddings exhibit varying degrees of precision sensitivity, which implies that a mixed-precision quantization strategy should be used that can be dynamically tuned at the granularity of individual features. The resulting storage savings can be strategically reinvested to enhance model capabilities through expanded sequence lengths and increased embedding dimensionality, ultimately achieving both infrastructure cost reduction and improved model performance in production environments.

Recommender Systems operate on three distinct feature types, each with specific data representations: 1) Dense features are characterized by continuous {\tt FLOAT} or {\tt DOUBLE} numerical values, include both direct measurements (e.g., demographic attributes, engagement metrics) and derived features generated through upstream offline model inference. 
2) Sparse features represent high-cardinality categorical variables, predominantly implemented through efficient lookup mechanisms such as {\tt INT} or {\tt BIGINT} foreign key references, where the feature space exhibits significant sparsity (e.g., user interaction histories, entity identifiers). 3) Embedding features are learned dense vector representations, typically encoded as {\tt FLOAT32} or {\tt FLOAT64} arrays, transformed from either dense or sparse inputs through parameterized embedding layers during model training.

Many existing technology companies face constraints from third-party search APIs (e.g., from Google or Bing) that motivate them to build their own proprietary search infrastructure for result aggregation and LLM-based summarization. These constraints include limitations on query volumes, restrictions on logging search results for model training, and stringent latency requirements. Consequently, they implement sophisticated crawling systems that operate at varying frequencies (minute/hour/day) based on domain priorities, while managing substantial computational resources for preprocessing and generating embeddings using offline ML models for similarity-based retrieval. 
Since the crawled data may include hundreds of trillions of pages, the storage of embeddings and associated metadata (such as entity attributes) poses significant capacity challenges, thereby increasing the motivation for vector embedding quantization techniques.

For integer features, quantization provides lossless compression by rehashing the input space to a smaller range (e.g., {\tt INT8}, {\tt INT16}, {\tt INT32}). For low cardinality columns, column stores can further leverage bit-packed encoding and RLE to achieve higher compression ratios. Native support for low-precision data types not only saves storage space, but also reduces network bandwidth and preprocessing computation overhead. 
Given that sparse features (integers) often constitute a significant portion of the total storage footprint in recommender systems, new columnar formats should incorporate nested light-weight adaptive encoding schemes~\cite{kuschewski2023btrblocks, nimble, chattopadhyay2019procella} to improve compression while minimizing decoding overhead. For float-point features and embeddings, 
the precision offered by 4-byte floats is often unnecessary for effective results. Feature quantization therefore stores values in {\tt FP16}, {\tt BF16}, or {\tt FP8} formats, usable directly in training and serving.
This reduction to 1 or 2 bytes per float can halve or quarter storage costs, disk I/O, network bandwidth, decompression efforts, and compute cost
during training and inference. 

{\bf Opportunities.} There are three potential areas that could improve the performance of the approach discussed above. First, native support for the reduced-precision formats ({\tt FP16}, {\tt BF16} and {\tt BF8}) in data processing frameworks (e.g., Apache Spark, Arrow, Velox) would enhance computational and memory efficiency during feature preprocessing (current implementations require interim solutions where data ingestion pipelines must implement automatic padding mechanisms to convert {\tt BF16} to {\tt FP32} format). Second, given that embedding vectors are typically normalized to (-1, 1), there exists a need for more storage-efficient encoding schemes specifically optimized for {\tt BF16}'s numerical characteristics. 
Third, some FP32 features are crucial for business-critical models. To mitigate potential accuracy degradation from FP16 quantization while maintaining computational efficiency, it is possible to use a dual-column storage strategy: decomposing FP32 features into two FP16 representations. This approach enables business-critical models to reconstruct original FP32 precision through 1:1 join operations during feature retrieval, while allowing other models to utilize FP16 features. Although this strategy does not reduce storage footprint, it  
decreases network bandwidth utilization and computational overhead in upstream processing stages.

\subsection{Multimodal Storage}

As stated in Section \ref{sec:intro}, LLM pre-training increasingly demands multi-modal data integration, encompassing text, images, audio, and video content. These data types typically appear in contextually related forms, such as within web pages, serving as collective inputs for model pre-training. Meanwhile, conventional column stores face challenges in supporting multi-modal data, 
especially for high-resolution video content. 

{\bf Challenge.} As described in Section \ref{sec:intro}, it is possible to use 
a dual-table architecture where meta tables leverage columnar format for efficient metadata management (including text, image and even audio), while media tables adopt row-oriented storage for multi-modal content. 
However, this approach
introduces a performance bottleneck during training. The separation of data across different storage formats (columnar for metadata and row-oriented for media) and locations
results in fragmented I/O operations, significantly impacting training I/O throughput and overall system performance. 

\begin{figure}
\centering
\includegraphics[width=0.45\textwidth]{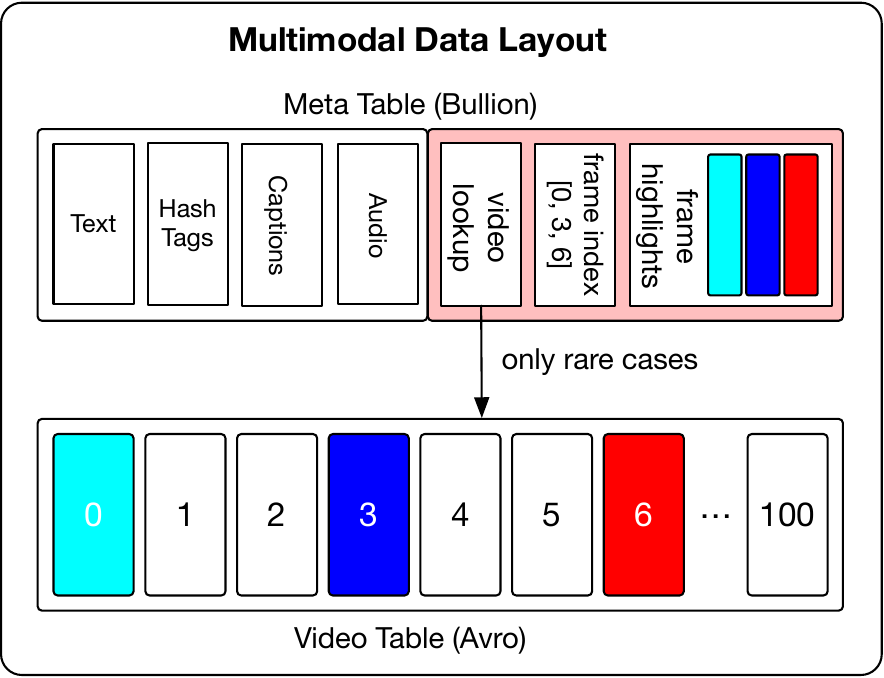}
\caption{Multimodal data layout.}
\label{metatable}
\end{figure}

{\bf Solution.} While specialized diffusion transformers like Sora~\cite{sora} require high-resolution, full-size videos for generating complex scenes with multiple characters, specific motion patterns, and accurate subject-background details, the pre-training of general-purpose LLMs does not require this same level of resolution. 
Instead, it can effectively operate on a strategic subset of video frames at reduced resolution, with these critical frames directly integrated into the columnar format, as shown in Figure~\ref{metatable}. This optimization enables the training system to access text, audio, and video modalities through the columnar storage alone, satisfying the majority of training scenarios. 
By having all the required data stored together, this approach eliminates the latency overhead associated with external, fragmented I/O operations. When full-size video access is required, the system still maintains the capability to perform external lookups through video indices stored in the meta table.

The meta table faces an additional challenge of random I/O patterns during training, as only high-quality data samples are selected for the training process. The quality-based filtering mechanism, primarily driven by quality scores, typically excludes a significant portion of the data, resulting in non-sequential access patterns within the columnar storage. To alleviate this I/O inefficiency, the storage system implements a quality-aware data organization strategy: incoming row data is presorted by quality score in descending order prior to insertion into the storage. This presorting approach improves contiguous access to high-quality video frames during training.

The quality-based reordering strategy for LLM training is a {\bf row reordering} approach, where entire rows are sorted based on their quality scores, ensuring that high-quality video frames and their associated features are stored contiguously within pages of each row group. This contrasts with the column-oriented reordering requirements in recommendation systems, where typically only 10\% of the thousands of available features are accessed during training. In recommendation scenarios, the system prioritizes frequently accessed, important features through {\bf column reordering}, ensuring these features (columns) are stored contiguously within row groups. 
Both approaches achieve their respective performance gains by minimizing random I/O through strategic data organization, though they operate on different dimensional axes of the storage structure.


\subsection{Cascading Encoding}

Given that machine learning workloads predominantly consist of integer and floating-point data types, adaptively selecting 
nested encoding schemes for each column at runtime can be highly effective. Recent work, including BtrBlocks~\cite{kuschewski2023btrblocks} and Nimble~\cite{nimble}, have adopted cascading encoding approaches;
however,
existing formats implement only a subset of available encoding schemes (listed   in Table~\ref{tab:encodings}), with no single implementation providing comprehensive coverage.
Furthermore, Parquet and ORC currently tightly couple various encoding methods (such as RLE, varint, and fixed-width encoding) without providing unified interfaces, making it impossible to utilize these encoding schemes independently. This coupling presents a significant barrier to implementing cascading encoding, which requires modular encoding selection.
What is needed is 
an independent encoding module
that provides cascading encoding capabilities. Such a modular approach would enable all columnar storage formats to leverage diverse encoding methods effectively.

\begin{table*}
\caption{Catalog of column encoding schemes
found in existing storage systems and formats including Parquet, ORC, Kudu, BtrBlocks~\cite{kuschewski2023btrblocks}, Nimble~\cite{nimble}, and Protocol Buffers.}
\label{tab:encodings}
\begin{tabular}{p{4cm}|p{10cm}}
\hline
\textbf{Encoding} & \textbf{Description} \\
\hline
Trival & A basic encoding scheme that stores data directly in its original format. \\
\hline
BitShuffle~\cite{bitshuffle} & A bit-level transformation that rearranges data by transposing a matrix of elements-by-bits, grouping bits of the same significance level together to improve compression efficiency. \\
\hline
RLE & Run-Length Encoding that compresses repeated values by storing distinct values and their consecutive occurrence counts in separate subcolumns. \\
\hline
Dictionary & Compresses data by maintaining a dictionary of unique values and storing data as indices referencing this dictionary. \\
\hline
FixedBitWidth & Compresses integer data using a uniform bit width for all values, optimized for cases with known value ranges. \\
\hline
Huffman & An entropy-based encoding optimized for integer values in the small range, assigning shorter codes to more frequent values. \\
\hline
Nullable & Handles null values using a two-subcolumn structure: one for null indicators and another for non-null values. \\
\hline
SparseBool & An optimized bitmap encoding for boolean values, typically used as a subcolumn in Nullable encoding for efficient null tracking. \\
\hline
Varint & Variable-length integer encoding that uses fewer bytes for smaller values, optimizing storage for integer distributions. \\
\hline
ZigZag & Encodes signed integers into unsigned values using zigzag pattern, efficiently handling both positive and negative numbers. \\
\hline
Delta & Stores differences between consecutive values using three subcolumns: base values, deltas, and delta indicators. Effective for monotonic or slowly-changing sequences. \\
\hline
SIMDFastPFOR~\cite{FastPFor} & SIMD-optimized implementation of PFOR (Patched Frame-of-Reference) compression, leveraging parallel processing for improved performance. \\
\hline
SIMDFastBP128~\cite{FastPFor} & SIMD-optimized binary packing compression that processes 128-bit segments in parallel for enhanced compression speed. \\
\hline
Constant & Optimizes storage for columns containing a single repeated value by storing only the constant value. \\
\hline
MainlyConstant & Optimizes columns dominated by a single value, storing the constant value, positions of exceptions, and their corresponding values. Also known as Frequency Encoding. \\
\hline
Sentinel & Represents null values by designating an unused value as a sentinel marker, encoding the data in a single subcolumn. \\
\hline
Chunked & Applies zstd compression to fixed-size chunks (256KB) of raw data, particularly effective for ML datasets with local patterns. \\
\hline
FSST~\cite{boncz2020fsst} & Fast Static Symbol Table compression that identifies and compresses both full string repetitions and common substrings, optimized for structured string data like URLs and emails. \\
\hline
Gorilla~\cite{pelkonen2015gorilla}/Chimp~\cite{liakos2022chimp} & An optimization of the Gorilla algorithm for floating-point compression, exploiting patterns in XOR'd values' leading and trailing zeros. \\

\hline
Pseudodecimal~\cite{kuschewski2023btrblocks} & Specialized encoding for floating-point values using decimal representation, optimizing for human-readable numeric patterns. \\
\hline
ALP~\cite{afroozeh2023alp} & An adaptive scheme that uses a strongly enhanced version of PseudoDecimals~\cite{kuschewski2023btrblocks} to losslessly encode doubles as integers if they originated as decimals, and otherwise uses vectorized compression of the doubles' front bits. \\

\hline
Roaring Bitmaps~\cite{roaring} & Advanced bitmap encoding that dynamically switches between different container types based on data density. \\
\hline
\end{tabular}
\end{table*}

The composable and recursive nature of encodings enables combinatorial patterns that can achieve superior data compression compared to static, single-encoding approaches. 
However, the search space for optimal encoding combinations grows significantly as the catalog expands, requiring systems like Procella~\cite{chattopadhyay2019procella} and BtrBlocks~\cite{kuschewski2023btrblocks} to employ sampling-based distribution analysis and heuristic approaches for encoding selection. The unbounded recursive potential raises an important question regarding the optimal depth of recursion. Current implementations, such as BtrBlocks, pragmatically limit recursion to one or two levels, but determining the ideal recursion depth still requires investigation. 

While prior work~\cite{zeng2023vldb} suggests that "formats should not apply general-purpose block compression by default", 
general-purpose compression methods (such as Chunked encoding in Table~\ref{tab:encodings}) continue to demonstrate practical value in specific scenarios. This is particularly evident in time-series data where many columns are rarely accessed, and in recommendation systems where only approximately 10\% of thousands of features are frequently queried—scenarios where block compression proves highly effective for storage optimization.



\section{Related Work}


Zeng et al.~\cite{zeng2023vldb} conducted an in-depth analysis of widely-used open formats for analytical DBMSs, such as Apache Parquet and Apache ORC. Their work highlights performance bottlenecks and overhead, particularly in the context of wide-table projection. These findings resonate with the challenges tackled by Bullion, underscoring the necessity for more efficient metadata management and feature projection mechanisms. 

Meta has unveiled Alpha~\cite{Chattopadhyay2023SharedFM,zhao2022understanding}, a novel columnar format aimed at boosting wide-table projection efficiency, especially for ML use cases, addressing significant weaknesses in previous formats. It includes: 1) feature flattening, which stores each feature as a separate stream on disk—effectively treating n features as n columns instead of a single map. 
2) feature reordering, which arranges commonly accessed features in adjacent disk positions to minimize over-reads from storage. 3) coalesced reads, which bundle selected feature streams into single I/O operations of up to 1.25 MiB, 
thereby amortizing disk seeks and improving throughput. 
Despite these advancements, Alpha does not explicitly address the metadata overhead challenge, data compliance, sparse feature encoding, and feature quantization, which are primary focuses of Bullion.

Recent advances in columnar storage have emerged from both academia and industry, 
including
BtrBlocks~\cite{kuschewski2023btrblocks} and Nimble~\cite{nimble}, which draw inspiration from Google's Procella~\cite{chattopadhyay2019procella}. These new formats revisit the concept of cascading/nested encodings at runtime, albeit with different optimization objectives. BtrBlocks primarily focuses on optimizing decoding efficiency through its nested encoding scheme, while Nimble introduces a more comprehensive and granular approach to encoding selection. Nimble incorporates a
user-configurable linear objective function that independently weights read time, write time, and storage size
that
enables users to tailor encoding strategies to their specific workload requirements.
These research efforts focus on improving data encoding, but do not comprehensively address the storage-level challenges and opportunities unique to machine learning workloads which is the focus of Bullion.

\section{Conclusion}

Bullion represents a step forward in the evolution of columnar storage systems, specifically designed to address the unique challenges and opportunities presented by the rapid growth of machine learning workloads. By introducing novel techniques such as hybrid deletion-compliance, optimized encoding for long sequence sparse features, feature quantization, efficient wide-table projection, quality-aware multimodal data organization, and a comprehensive cascading encoding framework, Bullion demonstrates the potential for columnar storage to adapt and excel in the era of AI and ML. As the demand for efficient and scalable data management solutions continues to grow, Bullion serves as a foundation for future research and development in this critical area.

\begin{acks}
We would like to express our gratitude to Yixin Wu, Kai Xie, Han Qian, and other members of the ByteDance Magnus data lake team for providing us with insights into the pain points and challenges encountered in storing machine learning training data. We also thank Hui Zhang, Yonghua Ding and Le Cai for their valuable participation in the early discussions.
\end{acks}

\bibliographystyle{ACM-Reference-Format}
\bibliography{sample-base}


\end{document}